\documentstyle[12pt]{article}
\begin{document}
\begin{center}
{\large {\bf Effects of boundary conditions on the critical}}
\end{center}
\begin{center}
{\large {\bf spanning probability}}
\end{center}
\bigskip
\begin{center} Muktish Acharyya$^+$ and Dietrich Stauffer$^*$ \end{center}
\vspace {0.2 cm}
\begin{center} {\it Institute for Theoretical Physics, University of Cologne}
\end{center}
\begin{center} {\it D-50923, Cologne, Germany} \end{center}

\bigskip
\bigskip
\bigskip
\begin{flushleft} {\bf Abstract:}\end{flushleft} The fractions of samples
spanning a lattice at its percolation threshold are 
found by computer simulation
of random site-percolation in two- and 
three-dimensional hypercubic lattices using
different boundary conditions. As a byproduct we find $p_c = 0.311605(5)$
in the cubic lattice.

\bigskip
\bigskip

\bigskip
\bigskip
\bigskip
\bigskip
\bigskip
\bigskip

\leftline {$^+$E-mail:muktish@thp.uni-koeln.de}
\leftline {$^*$E-mail:stauffer@thp.uni-koeln.de}
\newpage
\begin{flushleft}{\bf Introduction:}\end{flushleft}

In the case of random site percolation, the probability that the system
is percolating below the percolation threshold ($p_c$) is zero and it is unity
above $p_c$.
What is the probability of a cluster to span the lattice at the percolation
threshold \cite{sa}? 
The earlier studies [2-6] tried to
establish this quantity (spanning probability at percolation threshold) as
an universal quantity. 
To check the robustness of the spanning probability, we have studied here
the effect of different boundary conditions on the spanning probability at
percolation threshold. Our study is confined to two and three
dimensions and restricted to random site percolation. 

For free boundary conditions, in three dimensions,
the critical spanning probability
is $0.265\pm0.005$ and it is universal \cite{lhc}. However, an
earlier study \cite{saa} gives different result
($\sim 0.42$). 
This is due to the different boundary conditions used by
two groups of researchers as pointed out in [4,6]. 
Here, we study the critical spanning
probability (in two and three dimensions for random site percolation) and
its dependence on the different boundary condition.

For clarity, let us first define our boundary conditions: In D-dimensions
the spanning direction is always free

(1) Free boundary condition (FBC): 
The other D-1 directions are also free.
(2) Helical boundary condition (HBC): The Helical boundary conditions are
used in all D-1 remaining directions.
(3) Mixed boundary condition (MBC): This is not applicable in 2 dimensions.
In 3 dimensions, the Helical boundary 
condition is only applied in one of the directions of the planes and the
other direction is free.

We are considering only site percolation. The microscopic connectivity
is considered by the nearest neighbour occupied sites 
only and the macroscopic connectivity
(spanning) is checked by standard Hoshen-Kopelman \cite{sa} algorithm.
The spanning is checked only in one direction.

\bigskip

\begin{flushleft}{\bf Results:}\end{flushleft}

In $D = 2$ we have taken $p_c = 0.592746$ \cite{sa}. For a fixed system size
$L$ we occupied the lattice randomly with probability $p_c$. We have used
Kirkpatrick-Stoll R250 random number generator as well as multiplication
with 16807 \cite{sa}. 
Then we have checked whether the lattice is percolating
or not by standard Hoshen-Kopelman algorithm \cite{sa}
for $N_s$ number of different random samples. The number
(fraction) of samples percolating at $p_c$ is called the 
critical spanning
probability ($R_p$). For a particular boundary condition, the spanning
probability $R_p$  at $p_c$ 
has been calculated for different system sizes (for 
example, ranging
from $L=17$ to $L=6551$ in two dimensions). 
We have calculated the critical spanning probability in 2 and 3 dimensions
for the maximum allowable (in 128 MByte CRAY) system size.
The spanning probability, $R_p$  (in two dimensions) has been plotted
against $L$ for free boundary conditions (see Fig. 1). From the
extrapolation, we found $R_p = 0.501\pm0.003$. This is in good agreement
with the earlier result $(R_p = 1/2)$
\cite{ziff}. The
spanning probability for helical boundary condition in two dimensions was given
earlier \cite{hovi} as 0.637
and we report $R_p \sim 0.638$ in the asymptotic
limit ($L \to \infty$) (Fig.2).

In $D=3$, we have calculated the spanning probability $R_p$ at
  $p_c = 0.3116$ \cite{sa}. Here, also we have calculated $R_p$
for different boundary conditions in the same way described above. Fig. 3
shows the plots of $R_p$ against $L$ for free boundary conditions.
From extrapolation we obtained, $R_p \sim 0.28$, 
which is slightly higher than recently reported \cite{lhc} value. 
Using the mixed boundary condition, we have calculated the critical
spanning probability $R_p \sim 0.41$ in the asymptotic limit (Fig.4). This
result is slightly smaller than the earlier estimate \cite{saa}. In the
earlier case \cite{saa} the 
mixed boundary condition is used in the following way:
the top layer is fully occupied, the helical boundary condition is used in one
direction of the plane. But in the present case the top layer is randomly
occupied with probability $p_c$, and the helical bounadry condition is
used in one direction of the plane. These two slightly different boundary
conditions give different result in the smaller systems, however in the
$L \to \infty$ limit both boundary conditions give same result \cite{prv}.
In addition, in three dimensions, we have calculated the critical spanning
probability by using helical boundary condition in both directions of
the planes. Our estimated value of $R_p$ using helical boundary condition
is approximately $0.513$ (Fig.5). In each case, 
in three dimensions, we observed that
the value of $R_p$ is quite sensitive near $p_c$. For this reason we have
shown some results by taking $p_c = 0.31161$. All these results also indicate
that $p_c$ lies in between 0.3116 and 0.31161, compatible with earlier
estimates \cite{grp}.

In the present study, most of the data was obtained by taking $N_s = 10000$ in
each processor of a CRAY-T3E computer (using 64 processors).
The CPU time required, for the whole study,
can be calculated from the following 
example: for system size
$L = 481$ in three dimensions, each processor
averaged over
 $N_s=800$ different random realisations. The total CPU time
required for this is approximately 11 hours per processor.

\begin{flushleft}{\bf Summary:}\end{flushleft}

In summary, we studied the critical spanning probability in
two- and three- dimensional system using different boundary conditions. 
Our results from numerical
simulations (with larger system sizes and using better statistics)
show that the critical spanning probability is strongly dependent
on the boundary condition. 
There are different alternatives to helical boundary conditions in 3
dimensions, as pointed out in \cite{lhc}.
The value of the critical spanning probability
is quite sensitive to the value of $p_c$ used. This has been shown in three
dimensions and the size dependence of critical spanning probability can be
an alternative tool for estimating $p_c$.

\begin{flushleft}{\bf Acknowledgements:}\end{flushleft}
Let us thank R. M. Ziff for important discussions, SFB 341 for financial
support and HLRZ, J\"ulich for CRAY-T3E time.

\begin{flushleft} {\bf References:} \end{flushleft}

\newpage

% GNUPLOT: LaTeX picture
% Fig-1
\setlength{\unitlength}{0.240900pt}
\ifx\plotpoint\undefined\newsavebox{\plotpoint}\fi
\sbox{\plotpoint}{\rule[-0.200pt]{0.400pt}{0.400pt}}%
\begin{picture}(1500,900)(0,0)
\font\gnuplot=cmr10 at 10pt
\gnuplot
\sbox{\plotpoint}{\rule[-0.200pt]{0.400pt}{0.400pt}}%
\put(220.0,113.0){\rule[-0.200pt]{4.818pt}{0.400pt}}
\put(198,113){\makebox(0,0)[r]{0.48}}
\put(1416.0,113.0){\rule[-0.200pt]{4.818pt}{0.400pt}}
\put(220.0,209.0){\rule[-0.200pt]{4.818pt}{0.400pt}}
\put(198,209){\makebox(0,0)[r]{0.485}}
\put(1416.0,209.0){\rule[-0.200pt]{4.818pt}{0.400pt}}
\put(220.0,304.0){\rule[-0.200pt]{4.818pt}{0.400pt}}
\put(198,304){\makebox(0,0)[r]{0.49}}
\put(1416.0,304.0){\rule[-0.200pt]{4.818pt}{0.400pt}}
\put(220.0,400.0){\rule[-0.200pt]{4.818pt}{0.400pt}}
\put(198,400){\makebox(0,0)[r]{0.495}}
\put(1416.0,400.0){\rule[-0.200pt]{4.818pt}{0.400pt}}
\put(220.0,495.0){\rule[-0.200pt]{4.818pt}{0.400pt}}
\put(198,495){\makebox(0,0)[r]{0.5}}
\put(1416.0,495.0){\rule[-0.200pt]{4.818pt}{0.400pt}}
\put(220.0,591.0){\rule[-0.200pt]{4.818pt}{0.400pt}}
\put(198,591){\makebox(0,0)[r]{0.505}}
\put(1416.0,591.0){\rule[-0.200pt]{4.818pt}{0.400pt}}
\put(220.0,686.0){\rule[-0.200pt]{4.818pt}{0.400pt}}
\put(198,686){\makebox(0,0)[r]{0.51}}
\put(1416.0,686.0){\rule[-0.200pt]{4.818pt}{0.400pt}}
\put(220.0,782.0){\rule[-0.200pt]{4.818pt}{0.400pt}}
\put(198,782){\makebox(0,0)[r]{0.515}}
\put(1416.0,782.0){\rule[-0.200pt]{4.818pt}{0.400pt}}
\put(220.0,877.0){\rule[-0.200pt]{4.818pt}{0.400pt}}
\put(198,877){\makebox(0,0)[r]{0.52}}
\put(1416.0,877.0){\rule[-0.200pt]{4.818pt}{0.400pt}}
\put(220.0,113.0){\rule[-0.200pt]{0.400pt}{4.818pt}}
\put(220,68){\makebox(0,0){10}}
\put(220.0,857.0){\rule[-0.200pt]{0.400pt}{4.818pt}}
\put(342.0,113.0){\rule[-0.200pt]{0.400pt}{2.409pt}}
\put(342.0,867.0){\rule[-0.200pt]{0.400pt}{2.409pt}}
\put(413.0,113.0){\rule[-0.200pt]{0.400pt}{2.409pt}}
\put(413.0,867.0){\rule[-0.200pt]{0.400pt}{2.409pt}}
\put(464.0,113.0){\rule[-0.200pt]{0.400pt}{2.409pt}}
\put(464.0,867.0){\rule[-0.200pt]{0.400pt}{2.409pt}}
\put(503.0,113.0){\rule[-0.200pt]{0.400pt}{2.409pt}}
\put(503.0,867.0){\rule[-0.200pt]{0.400pt}{2.409pt}}
\put(535.0,113.0){\rule[-0.200pt]{0.400pt}{2.409pt}}
\put(535.0,867.0){\rule[-0.200pt]{0.400pt}{2.409pt}}
\put(563.0,113.0){\rule[-0.200pt]{0.400pt}{2.409pt}}
\put(563.0,867.0){\rule[-0.200pt]{0.400pt}{2.409pt}}
\put(586.0,113.0){\rule[-0.200pt]{0.400pt}{2.409pt}}
\put(586.0,867.0){\rule[-0.200pt]{0.400pt}{2.409pt}}
\put(607.0,113.0){\rule[-0.200pt]{0.400pt}{2.409pt}}
\put(607.0,867.0){\rule[-0.200pt]{0.400pt}{2.409pt}}
\put(625.0,113.0){\rule[-0.200pt]{0.400pt}{4.818pt}}
\put(625,68){\makebox(0,0){100}}
\put(625.0,857.0){\rule[-0.200pt]{0.400pt}{4.818pt}}
\put(747.0,113.0){\rule[-0.200pt]{0.400pt}{2.409pt}}
\put(747.0,867.0){\rule[-0.200pt]{0.400pt}{2.409pt}}
\put(819.0,113.0){\rule[-0.200pt]{0.400pt}{2.409pt}}
\put(819.0,867.0){\rule[-0.200pt]{0.400pt}{2.409pt}}
\put(869.0,113.0){\rule[-0.200pt]{0.400pt}{2.409pt}}
\put(869.0,867.0){\rule[-0.200pt]{0.400pt}{2.409pt}}
\put(909.0,113.0){\rule[-0.200pt]{0.400pt}{2.409pt}}
\put(909.0,867.0){\rule[-0.200pt]{0.400pt}{2.409pt}}
\put(941.0,113.0){\rule[-0.200pt]{0.400pt}{2.409pt}}
\put(941.0,867.0){\rule[-0.200pt]{0.400pt}{2.409pt}}
\put(968.0,113.0){\rule[-0.200pt]{0.400pt}{2.409pt}}
\put(968.0,867.0){\rule[-0.200pt]{0.400pt}{2.409pt}}
\put(991.0,113.0){\rule[-0.200pt]{0.400pt}{2.409pt}}
\put(991.0,867.0){\rule[-0.200pt]{0.400pt}{2.409pt}}
\put(1012.0,113.0){\rule[-0.200pt]{0.400pt}{2.409pt}}
\put(1012.0,867.0){\rule[-0.200pt]{0.400pt}{2.409pt}}
\put(1031.0,113.0){\rule[-0.200pt]{0.400pt}{4.818pt}}
\put(1031,68){\makebox(0,0){1000}}
\put(1031.0,857.0){\rule[-0.200pt]{0.400pt}{4.818pt}}
\put(1153.0,113.0){\rule[-0.200pt]{0.400pt}{2.409pt}}
\put(1153.0,867.0){\rule[-0.200pt]{0.400pt}{2.409pt}}
\put(1224.0,113.0){\rule[-0.200pt]{0.400pt}{2.409pt}}
\put(1224.0,867.0){\rule[-0.200pt]{0.400pt}{2.409pt}}
\put(1275.0,113.0){\rule[-0.200pt]{0.400pt}{2.409pt}}
\put(1275.0,867.0){\rule[-0.200pt]{0.400pt}{2.409pt}}
\put(1314.0,113.0){\rule[-0.200pt]{0.400pt}{2.409pt}}
\put(1314.0,867.0){\rule[-0.200pt]{0.400pt}{2.409pt}}
\put(1346.0,113.0){\rule[-0.200pt]{0.400pt}{2.409pt}}
\put(1346.0,867.0){\rule[-0.200pt]{0.400pt}{2.409pt}}
\put(1373.0,113.0){\rule[-0.200pt]{0.400pt}{2.409pt}}
\put(1373.0,867.0){\rule[-0.200pt]{0.400pt}{2.409pt}}
\put(1397.0,113.0){\rule[-0.200pt]{0.400pt}{2.409pt}}
\put(1397.0,867.0){\rule[-0.200pt]{0.400pt}{2.409pt}}
\put(1417.0,113.0){\rule[-0.200pt]{0.400pt}{2.409pt}}
\put(1417.0,867.0){\rule[-0.200pt]{0.400pt}{2.409pt}}
\put(1436.0,113.0){\rule[-0.200pt]{0.400pt}{4.818pt}}
\put(1436,68){\makebox(0,0){10000}}
\put(1436.0,857.0){\rule[-0.200pt]{0.400pt}{4.818pt}}
\put(220.0,113.0){\rule[-0.200pt]{292.934pt}{0.400pt}}
\put(1436.0,113.0){\rule[-0.200pt]{0.400pt}{184.048pt}}
\put(220.0,877.0){\rule[-0.200pt]{292.934pt}{0.400pt}}
\put(45,495){\makebox(0,0){$R_p(p_c)$}}
\put(828,23){\makebox(0,0){$L$}}
\put(1224,782){\makebox(0,0)[l]{$D = 2$}}
\put(1224,686){\makebox(0,0)[l]{$FBC$}}
\put(220.0,113.0){\rule[-0.200pt]{0.400pt}{184.048pt}}
\sbox{\plotpoint}{\rule[-0.500pt]{1.000pt}{1.000pt}}%
\put(1436,495){\usebox{\plotpoint}}
\multiput(1436,495)(-20.756,0.000){20}{\usebox{\plotpoint}}
\multiput(1030,495)(-20.754,0.258){8}{\usebox{\plotpoint}}
\multiput(869,497)(-20.742,0.741){5}{\usebox{\plotpoint}}
\multiput(757,501)(-20.171,4.890){7}{\usebox{\plotpoint}}
\multiput(625,533)(-16.525,12.559){3}{\usebox{\plotpoint}}
\multiput(575,571)(-17.078,11.796){11}{\usebox{\plotpoint}}
\multiput(381,705)(-15.953,13.278){10}{\usebox{\plotpoint}}
\put(220,839){\usebox{\plotpoint}}
\sbox{\plotpoint}{\rule[-0.200pt]{0.400pt}{0.400pt}}%
\put(313,799){\makebox(0,0){$\bullet$}}
\put(367,717){\makebox(0,0){$\bullet$}}
\put(507,573){\makebox(0,0){$\bullet$}}
\put(584,586){\makebox(0,0){$\bullet$}}
\put(662,483){\makebox(0,0){$\bullet$}}
\put(757,501){\makebox(0,0){$\bullet$}}
\put(934,524){\makebox(0,0){$\bullet$}}
\put(1049,521){\makebox(0,0){$\bullet$}}
\put(1179,499){\makebox(0,0){$\bullet$}}
\put(1318,512){\makebox(0,0){$\bullet$}}
\put(313.0,774.0){\rule[-0.200pt]{0.400pt}{12.045pt}}
\put(303.0,774.0){\rule[-0.200pt]{4.818pt}{0.400pt}}
\put(303.0,824.0){\rule[-0.200pt]{4.818pt}{0.400pt}}
\put(367.0,698.0){\rule[-0.200pt]{0.400pt}{9.395pt}}
\put(357.0,698.0){\rule[-0.200pt]{4.818pt}{0.400pt}}
\put(357.0,737.0){\rule[-0.200pt]{4.818pt}{0.400pt}}
\put(507.0,556.0){\rule[-0.200pt]{0.400pt}{7.950pt}}
\put(497.0,556.0){\rule[-0.200pt]{4.818pt}{0.400pt}}
\put(497.0,589.0){\rule[-0.200pt]{4.818pt}{0.400pt}}
\put(584.0,556.0){\rule[-0.200pt]{0.400pt}{14.695pt}}
\put(574.0,556.0){\rule[-0.200pt]{4.818pt}{0.400pt}}
\put(574.0,617.0){\rule[-0.200pt]{4.818pt}{0.400pt}}
\put(662.0,449.0){\rule[-0.200pt]{0.400pt}{16.381pt}}
\put(652.0,449.0){\rule[-0.200pt]{4.818pt}{0.400pt}}
\put(652.0,517.0){\rule[-0.200pt]{4.818pt}{0.400pt}}
\put(757.0,471.0){\rule[-0.200pt]{0.400pt}{14.213pt}}
\put(747.0,471.0){\rule[-0.200pt]{4.818pt}{0.400pt}}
\put(747.0,530.0){\rule[-0.200pt]{4.818pt}{0.400pt}}
\put(934.0,502.0){\rule[-0.200pt]{0.400pt}{10.840pt}}
\put(924.0,502.0){\rule[-0.200pt]{4.818pt}{0.400pt}}
\put(924.0,547.0){\rule[-0.200pt]{4.818pt}{0.400pt}}
\put(1049.0,506.0){\rule[-0.200pt]{0.400pt}{7.227pt}}
\put(1039.0,506.0){\rule[-0.200pt]{4.818pt}{0.400pt}}
\put(1039.0,536.0){\rule[-0.200pt]{4.818pt}{0.400pt}}
\put(1179.0,487.0){\rule[-0.200pt]{0.400pt}{6.022pt}}
\put(1169.0,487.0){\rule[-0.200pt]{4.818pt}{0.400pt}}
\put(1169.0,512.0){\rule[-0.200pt]{4.818pt}{0.400pt}}
\put(1318.0,490.0){\rule[-0.200pt]{0.400pt}{10.840pt}}
\put(1308.0,490.0){\rule[-0.200pt]{4.818pt}{0.400pt}}
\put(1308.0,535.0){\rule[-0.200pt]{4.818pt}{0.400pt}}
\end{picture}

\bigskip

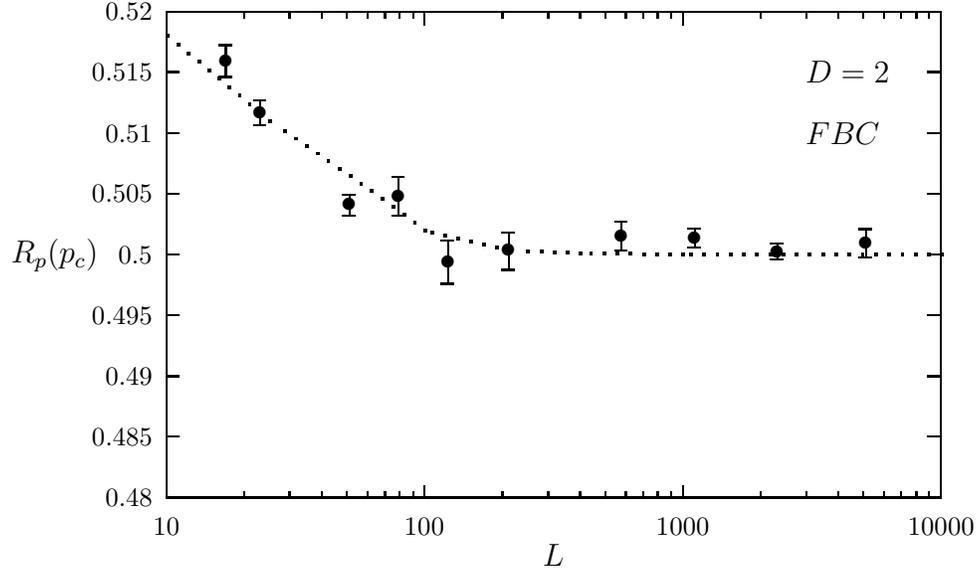
\noindent {\bf Fig. 1.} Size dependence of critical spanning probability
(using $p_c = 0.592746$) in two dimensions for free boundary conditions. 
Dotted line is a guide to the eye.

\vspace {0.3 cm}

% GNUPLOT: LaTeX picture
% Fig-2
\setlength{\unitlength}{0.240900pt}
\ifx\plotpoint\undefined\newsavebox{\plotpoint}\fi
\sbox{\plotpoint}{\rule[-0.200pt]{0.400pt}{0.400pt}}%
\begin{picture}(1500,900)(0,0)
\font\gnuplot=cmr10 at 10pt
\gnuplot
\sbox{\plotpoint}{\rule[-0.200pt]{0.400pt}{0.400pt}}%
\put(220.0,113.0){\rule[-0.200pt]{4.818pt}{0.400pt}}
\put(198,113){\makebox(0,0)[r]{0.61}}
\put(1416.0,113.0){\rule[-0.200pt]{4.818pt}{0.400pt}}
\put(220.0,209.0){\rule[-0.200pt]{4.818pt}{0.400pt}}
\put(198,209){\makebox(0,0)[r]{0.615}}
\put(1416.0,209.0){\rule[-0.200pt]{4.818pt}{0.400pt}}
\put(220.0,304.0){\rule[-0.200pt]{4.818pt}{0.400pt}}
\put(198,304){\makebox(0,0)[r]{0.62}}
\put(1416.0,304.0){\rule[-0.200pt]{4.818pt}{0.400pt}}
\put(220.0,400.0){\rule[-0.200pt]{4.818pt}{0.400pt}}
\put(198,400){\makebox(0,0)[r]{0.625}}
\put(1416.0,400.0){\rule[-0.200pt]{4.818pt}{0.400pt}}
\put(220.0,495.0){\rule[-0.200pt]{4.818pt}{0.400pt}}
\put(198,495){\makebox(0,0)[r]{0.63}}
\put(1416.0,495.0){\rule[-0.200pt]{4.818pt}{0.400pt}}
\put(220.0,591.0){\rule[-0.200pt]{4.818pt}{0.400pt}}
\put(198,591){\makebox(0,0)[r]{0.635}}
\put(1416.0,591.0){\rule[-0.200pt]{4.818pt}{0.400pt}}
\put(220.0,686.0){\rule[-0.200pt]{4.818pt}{0.400pt}}
\put(198,686){\makebox(0,0)[r]{0.64}}
\put(1416.0,686.0){\rule[-0.200pt]{4.818pt}{0.400pt}}
\put(220.0,782.0){\rule[-0.200pt]{4.818pt}{0.400pt}}
\put(198,782){\makebox(0,0)[r]{0.645}}
\put(1416.0,782.0){\rule[-0.200pt]{4.818pt}{0.400pt}}
\put(220.0,877.0){\rule[-0.200pt]{4.818pt}{0.400pt}}
\put(198,877){\makebox(0,0)[r]{0.65}}
\put(1416.0,877.0){\rule[-0.200pt]{4.818pt}{0.400pt}}
\put(220.0,113.0){\rule[-0.200pt]{0.400pt}{4.818pt}}
\put(220,68){\makebox(0,0){10}}
\put(220.0,857.0){\rule[-0.200pt]{0.400pt}{4.818pt}}
\put(342.0,113.0){\rule[-0.200pt]{0.400pt}{2.409pt}}
\put(342.0,867.0){\rule[-0.200pt]{0.400pt}{2.409pt}}
\put(413.0,113.0){\rule[-0.200pt]{0.400pt}{2.409pt}}
\put(413.0,867.0){\rule[-0.200pt]{0.400pt}{2.409pt}}
\put(464.0,113.0){\rule[-0.200pt]{0.400pt}{2.409pt}}
\put(464.0,867.0){\rule[-0.200pt]{0.400pt}{2.409pt}}
\put(503.0,113.0){\rule[-0.200pt]{0.400pt}{2.409pt}}
\put(503.0,867.0){\rule[-0.200pt]{0.400pt}{2.409pt}}
\put(535.0,113.0){\rule[-0.200pt]{0.400pt}{2.409pt}}
\put(535.0,867.0){\rule[-0.200pt]{0.400pt}{2.409pt}}
\put(563.0,113.0){\rule[-0.200pt]{0.400pt}{2.409pt}}
\put(563.0,867.0){\rule[-0.200pt]{0.400pt}{2.409pt}}
\put(586.0,113.0){\rule[-0.200pt]{0.400pt}{2.409pt}}
\put(586.0,867.0){\rule[-0.200pt]{0.400pt}{2.409pt}}
\put(607.0,113.0){\rule[-0.200pt]{0.400pt}{2.409pt}}
\put(607.0,867.0){\rule[-0.200pt]{0.400pt}{2.409pt}}
\put(625.0,113.0){\rule[-0.200pt]{0.400pt}{4.818pt}}
\put(625,68){\makebox(0,0){100}}
\put(625.0,857.0){\rule[-0.200pt]{0.400pt}{4.818pt}}
\put(747.0,113.0){\rule[-0.200pt]{0.400pt}{2.409pt}}
\put(747.0,867.0){\rule[-0.200pt]{0.400pt}{2.409pt}}
\put(819.0,113.0){\rule[-0.200pt]{0.400pt}{2.409pt}}
\put(819.0,867.0){\rule[-0.200pt]{0.400pt}{2.409pt}}
\put(869.0,113.0){\rule[-0.200pt]{0.400pt}{2.409pt}}
\put(869.0,867.0){\rule[-0.200pt]{0.400pt}{2.409pt}}
\put(909.0,113.0){\rule[-0.200pt]{0.400pt}{2.409pt}}
\put(909.0,867.0){\rule[-0.200pt]{0.400pt}{2.409pt}}
\put(941.0,113.0){\rule[-0.200pt]{0.400pt}{2.409pt}}
\put(941.0,867.0){\rule[-0.200pt]{0.400pt}{2.409pt}}
\put(968.0,113.0){\rule[-0.200pt]{0.400pt}{2.409pt}}
\put(968.0,867.0){\rule[-0.200pt]{0.400pt}{2.409pt}}
\put(991.0,113.0){\rule[-0.200pt]{0.400pt}{2.409pt}}
\put(991.0,867.0){\rule[-0.200pt]{0.400pt}{2.409pt}}
\put(1012.0,113.0){\rule[-0.200pt]{0.400pt}{2.409pt}}
\put(1012.0,867.0){\rule[-0.200pt]{0.400pt}{2.409pt}}
\put(1031.0,113.0){\rule[-0.200pt]{0.400pt}{4.818pt}}
\put(1031,68){\makebox(0,0){1000}}
\put(1031.0,857.0){\rule[-0.200pt]{0.400pt}{4.818pt}}
\put(1153.0,113.0){\rule[-0.200pt]{0.400pt}{2.409pt}}
\put(1153.0,867.0){\rule[-0.200pt]{0.400pt}{2.409pt}}
\put(1224.0,113.0){\rule[-0.200pt]{0.400pt}{2.409pt}}
\put(1224.0,867.0){\rule[-0.200pt]{0.400pt}{2.409pt}}
\put(1275.0,113.0){\rule[-0.200pt]{0.400pt}{2.409pt}}
\put(1275.0,867.0){\rule[-0.200pt]{0.400pt}{2.409pt}}
\put(1314.0,113.0){\rule[-0.200pt]{0.400pt}{2.409pt}}
\put(1314.0,867.0){\rule[-0.200pt]{0.400pt}{2.409pt}}
\put(1346.0,113.0){\rule[-0.200pt]{0.400pt}{2.409pt}}
\put(1346.0,867.0){\rule[-0.200pt]{0.400pt}{2.409pt}}
\put(1373.0,113.0){\rule[-0.200pt]{0.400pt}{2.409pt}}
\put(1373.0,867.0){\rule[-0.200pt]{0.400pt}{2.409pt}}
\put(1397.0,113.0){\rule[-0.200pt]{0.400pt}{2.409pt}}
\put(1397.0,867.0){\rule[-0.200pt]{0.400pt}{2.409pt}}
\put(1417.0,113.0){\rule[-0.200pt]{0.400pt}{2.409pt}}
\put(1417.0,867.0){\rule[-0.200pt]{0.400pt}{2.409pt}}
\put(1436.0,113.0){\rule[-0.200pt]{0.400pt}{4.818pt}}
\put(1436,68){\makebox(0,0){10000}}
\put(1436.0,857.0){\rule[-0.200pt]{0.400pt}{4.818pt}}
\put(220.0,113.0){\rule[-0.200pt]{292.934pt}{0.400pt}}
\put(1436.0,113.0){\rule[-0.200pt]{0.400pt}{184.048pt}}
\put(220.0,877.0){\rule[-0.200pt]{292.934pt}{0.400pt}}
\put(45,495){\makebox(0,0){$R_p(p_c)$}}
\put(828,23){\makebox(0,0){$L$}}
\put(413,782){\makebox(0,0)[l]{$D = 2$}}
\put(413,686){\makebox(0,0)[l]{$HBC$}}
\put(220.0,113.0){\rule[-0.200pt]{0.400pt}{184.048pt}}
\sbox{\plotpoint}{\rule[-0.500pt]{1.000pt}{1.000pt}}%
\put(1436,648){\usebox{\plotpoint}}
\multiput(1436,648)(-20.752,-0.392){11}{\usebox{\plotpoint}}
\multiput(1224,644)(-20.656,-2.033){9}{\usebox{\plotpoint}}
\multiput(1031,625)(-20.473,-3.412){4}{\usebox{\plotpoint}}
\multiput(941,610)(-20.304,-4.304){10}{\usebox{\plotpoint}}
\multiput(757,571)(-20.364,-4.011){6}{\usebox{\plotpoint}}
\multiput(625,545)(-20.182,-4.844){2}{\usebox{\plotpoint}}
\multiput(575,533)(-19.637,-6.722){6}{\usebox{\plotpoint}}
\multiput(464,495)(-16.643,-12.401){3}{\usebox{\plotpoint}}
\multiput(413,457)(-13.369,-15.876){3}{\usebox{\plotpoint}}
\multiput(381,419)(-14.009,-15.314){11}{\usebox{\plotpoint}}
\put(220,243){\usebox{\plotpoint}}
\sbox{\plotpoint}{\rule[-0.200pt]{0.400pt}{0.400pt}}%
\put(313,370){\makebox(0,0){$\bullet$}}
\put(367,408){\makebox(0,0){$\bullet$}}
\put(507,530){\makebox(0,0){$\bullet$}}
\put(584,541){\makebox(0,0){$\bullet$}}
\put(662,536){\makebox(0,0){$\bullet$}}
\put(757,574){\makebox(0,0){$\bullet$}}
\put(934,627){\makebox(0,0){$\bullet$}}
\put(1049,627){\makebox(0,0){$\bullet$}}
\put(1179,625){\makebox(0,0){$\bullet$}}
\put(1318,636){\makebox(0,0){$\bullet$}}
\put(1362,675){\makebox(0,0){$\bullet$}}
\put(313.0,347.0){\rule[-0.200pt]{0.400pt}{11.081pt}}
\put(303.0,347.0){\rule[-0.200pt]{4.818pt}{0.400pt}}
\put(303.0,393.0){\rule[-0.200pt]{4.818pt}{0.400pt}}
\put(367.0,387.0){\rule[-0.200pt]{0.400pt}{10.118pt}}
\put(357.0,387.0){\rule[-0.200pt]{4.818pt}{0.400pt}}
\put(357.0,429.0){\rule[-0.200pt]{4.818pt}{0.400pt}}
\put(507.0,507.0){\rule[-0.200pt]{0.400pt}{10.840pt}}
\put(497.0,507.0){\rule[-0.200pt]{4.818pt}{0.400pt}}
\put(497.0,552.0){\rule[-0.200pt]{4.818pt}{0.400pt}}
\put(584.0,518.0){\rule[-0.200pt]{0.400pt}{11.081pt}}
\put(574.0,518.0){\rule[-0.200pt]{4.818pt}{0.400pt}}
\put(574.0,564.0){\rule[-0.200pt]{4.818pt}{0.400pt}}
\put(662.0,502.0){\rule[-0.200pt]{0.400pt}{16.622pt}}
\put(652.0,502.0){\rule[-0.200pt]{4.818pt}{0.400pt}}
\put(652.0,571.0){\rule[-0.200pt]{4.818pt}{0.400pt}}
\put(757.0,557.0){\rule[-0.200pt]{0.400pt}{8.191pt}}
\put(747.0,557.0){\rule[-0.200pt]{4.818pt}{0.400pt}}
\put(747.0,591.0){\rule[-0.200pt]{4.818pt}{0.400pt}}
\put(934.0,599.0){\rule[-0.200pt]{0.400pt}{13.249pt}}
\put(924.0,599.0){\rule[-0.200pt]{4.818pt}{0.400pt}}
\put(924.0,654.0){\rule[-0.200pt]{4.818pt}{0.400pt}}
\put(1049.0,613.0){\rule[-0.200pt]{0.400pt}{6.745pt}}
\put(1039.0,613.0){\rule[-0.200pt]{4.818pt}{0.400pt}}
\put(1039.0,641.0){\rule[-0.200pt]{4.818pt}{0.400pt}}
\put(1179.0,613.0){\rule[-0.200pt]{0.400pt}{5.541pt}}
\put(1169.0,613.0){\rule[-0.200pt]{4.818pt}{0.400pt}}
\put(1169.0,636.0){\rule[-0.200pt]{4.818pt}{0.400pt}}
\put(1318.0,615.0){\rule[-0.200pt]{0.400pt}{10.359pt}}
\put(1308.0,615.0){\rule[-0.200pt]{4.818pt}{0.400pt}}
\put(1308.0,658.0){\rule[-0.200pt]{4.818pt}{0.400pt}}
\put(1362.0,653.0){\rule[-0.200pt]{0.400pt}{10.600pt}}
\put(1352.0,653.0){\rule[-0.200pt]{4.818pt}{0.400pt}}
\put(1352.0,697.0){\rule[-0.200pt]{4.818pt}{0.400pt}}
\end{picture}

\bigskip

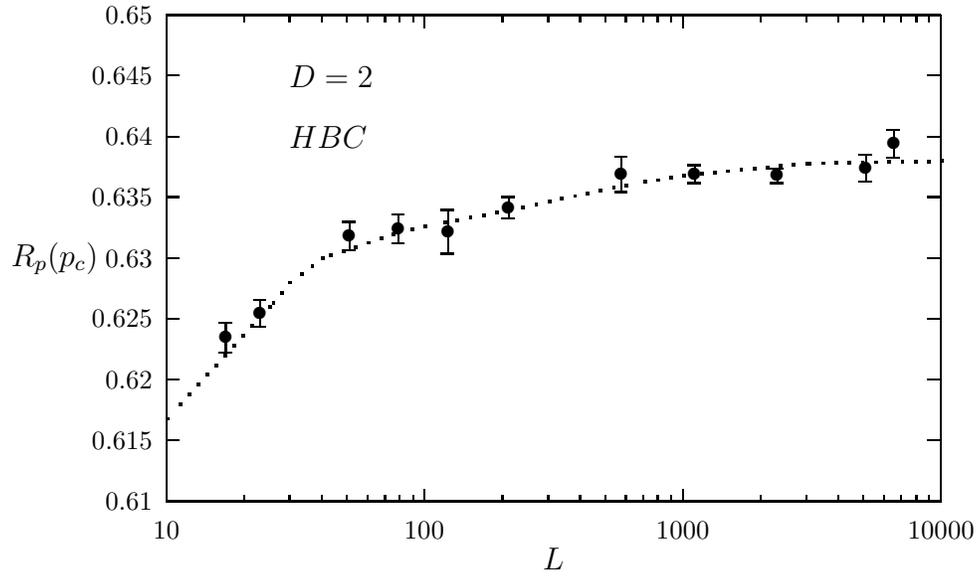
\noindent {\bf Fig. 2.} As Fig. 1, for helical boundary conditions.

\newpage

% GNUPLOT: LaTeX picture
% Fig-3
\setlength{\unitlength}{0.240900pt}
\ifx\plotpoint\undefined\newsavebox{\plotpoint}\fi
\sbox{\plotpoint}{\rule[-0.200pt]{0.400pt}{0.400pt}}%
\begin{picture}(1500,900)(0,0)
\font\gnuplot=cmr10 at 10pt
\gnuplot
\sbox{\plotpoint}{\rule[-0.200pt]{0.400pt}{0.400pt}}%
\put(220.0,113.0){\rule[-0.200pt]{4.818pt}{0.400pt}}
\put(198,113){\makebox(0,0)[r]{0.26}}
\put(1416.0,113.0){\rule[-0.200pt]{4.818pt}{0.400pt}}
\put(220.0,240.0){\rule[-0.200pt]{4.818pt}{0.400pt}}
\put(198,240){\makebox(0,0)[r]{0.27}}
\put(1416.0,240.0){\rule[-0.200pt]{4.818pt}{0.400pt}}
\put(220.0,368.0){\rule[-0.200pt]{4.818pt}{0.400pt}}
\put(198,368){\makebox(0,0)[r]{0.28}}
\put(1416.0,368.0){\rule[-0.200pt]{4.818pt}{0.400pt}}
\put(220.0,495.0){\rule[-0.200pt]{4.818pt}{0.400pt}}
\put(198,495){\makebox(0,0)[r]{0.29}}
\put(1416.0,495.0){\rule[-0.200pt]{4.818pt}{0.400pt}}
\put(220.0,622.0){\rule[-0.200pt]{4.818pt}{0.400pt}}
\put(198,622){\makebox(0,0)[r]{0.3}}
\put(1416.0,622.0){\rule[-0.200pt]{4.818pt}{0.400pt}}
\put(220.0,750.0){\rule[-0.200pt]{4.818pt}{0.400pt}}
\put(198,750){\makebox(0,0)[r]{0.31}}
\put(1416.0,750.0){\rule[-0.200pt]{4.818pt}{0.400pt}}
\put(220.0,877.0){\rule[-0.200pt]{4.818pt}{0.400pt}}
\put(198,877){\makebox(0,0)[r]{0.32}}
\put(1416.0,877.0){\rule[-0.200pt]{4.818pt}{0.400pt}}
\put(220.0,113.0){\rule[-0.200pt]{0.400pt}{4.818pt}}
\put(220,68){\makebox(0,0){1}}
\put(220.0,857.0){\rule[-0.200pt]{0.400pt}{4.818pt}}
\put(342.0,113.0){\rule[-0.200pt]{0.400pt}{2.409pt}}
\put(342.0,867.0){\rule[-0.200pt]{0.400pt}{2.409pt}}
\put(413.0,113.0){\rule[-0.200pt]{0.400pt}{2.409pt}}
\put(413.0,867.0){\rule[-0.200pt]{0.400pt}{2.409pt}}
\put(464.0,113.0){\rule[-0.200pt]{0.400pt}{2.409pt}}
\put(464.0,867.0){\rule[-0.200pt]{0.400pt}{2.409pt}}
\put(503.0,113.0){\rule[-0.200pt]{0.400pt}{2.409pt}}
\put(503.0,867.0){\rule[-0.200pt]{0.400pt}{2.409pt}}
\put(535.0,113.0){\rule[-0.200pt]{0.400pt}{2.409pt}}
\put(535.0,867.0){\rule[-0.200pt]{0.400pt}{2.409pt}}
\put(563.0,113.0){\rule[-0.200pt]{0.400pt}{2.409pt}}
\put(563.0,867.0){\rule[-0.200pt]{0.400pt}{2.409pt}}
\put(586.0,113.0){\rule[-0.200pt]{0.400pt}{2.409pt}}
\put(586.0,867.0){\rule[-0.200pt]{0.400pt}{2.409pt}}
\put(607.0,113.0){\rule[-0.200pt]{0.400pt}{2.409pt}}
\put(607.0,867.0){\rule[-0.200pt]{0.400pt}{2.409pt}}
\put(625.0,113.0){\rule[-0.200pt]{0.400pt}{4.818pt}}
\put(625,68){\makebox(0,0){10}}
\put(625.0,857.0){\rule[-0.200pt]{0.400pt}{4.818pt}}
\put(747.0,113.0){\rule[-0.200pt]{0.400pt}{2.409pt}}
\put(747.0,867.0){\rule[-0.200pt]{0.400pt}{2.409pt}}
\put(819.0,113.0){\rule[-0.200pt]{0.400pt}{2.409pt}}
\put(819.0,867.0){\rule[-0.200pt]{0.400pt}{2.409pt}}
\put(869.0,113.0){\rule[-0.200pt]{0.400pt}{2.409pt}}
\put(869.0,867.0){\rule[-0.200pt]{0.400pt}{2.409pt}}
\put(909.0,113.0){\rule[-0.200pt]{0.400pt}{2.409pt}}
\put(909.0,867.0){\rule[-0.200pt]{0.400pt}{2.409pt}}
\put(941.0,113.0){\rule[-0.200pt]{0.400pt}{2.409pt}}
\put(941.0,867.0){\rule[-0.200pt]{0.400pt}{2.409pt}}
\put(968.0,113.0){\rule[-0.200pt]{0.400pt}{2.409pt}}
\put(968.0,867.0){\rule[-0.200pt]{0.400pt}{2.409pt}}
\put(991.0,113.0){\rule[-0.200pt]{0.400pt}{2.409pt}}
\put(991.0,867.0){\rule[-0.200pt]{0.400pt}{2.409pt}}
\put(1012.0,113.0){\rule[-0.200pt]{0.400pt}{2.409pt}}
\put(1012.0,867.0){\rule[-0.200pt]{0.400pt}{2.409pt}}
\put(1031.0,113.0){\rule[-0.200pt]{0.400pt}{4.818pt}}
\put(1031,68){\makebox(0,0){100}}
\put(1031.0,857.0){\rule[-0.200pt]{0.400pt}{4.818pt}}
\put(1153.0,113.0){\rule[-0.200pt]{0.400pt}{2.409pt}}
\put(1153.0,867.0){\rule[-0.200pt]{0.400pt}{2.409pt}}
\put(1224.0,113.0){\rule[-0.200pt]{0.400pt}{2.409pt}}
\put(1224.0,867.0){\rule[-0.200pt]{0.400pt}{2.409pt}}
\put(1275.0,113.0){\rule[-0.200pt]{0.400pt}{2.409pt}}
\put(1275.0,867.0){\rule[-0.200pt]{0.400pt}{2.409pt}}
\put(1314.0,113.0){\rule[-0.200pt]{0.400pt}{2.409pt}}
\put(1314.0,867.0){\rule[-0.200pt]{0.400pt}{2.409pt}}
\put(1346.0,113.0){\rule[-0.200pt]{0.400pt}{2.409pt}}
\put(1346.0,867.0){\rule[-0.200pt]{0.400pt}{2.409pt}}
\put(1373.0,113.0){\rule[-0.200pt]{0.400pt}{2.409pt}}
\put(1373.0,867.0){\rule[-0.200pt]{0.400pt}{2.409pt}}
\put(1397.0,113.0){\rule[-0.200pt]{0.400pt}{2.409pt}}
\put(1397.0,867.0){\rule[-0.200pt]{0.400pt}{2.409pt}}
\put(1417.0,113.0){\rule[-0.200pt]{0.400pt}{2.409pt}}
\put(1417.0,867.0){\rule[-0.200pt]{0.400pt}{2.409pt}}
\put(1436.0,113.0){\rule[-0.200pt]{0.400pt}{4.818pt}}
\put(1436,68){\makebox(0,0){1000}}
\put(1436.0,857.0){\rule[-0.200pt]{0.400pt}{4.818pt}}
\put(220.0,113.0){\rule[-0.200pt]{292.934pt}{0.400pt}}
\put(1436.0,113.0){\rule[-0.200pt]{0.400pt}{184.048pt}}
\put(220.0,877.0){\rule[-0.200pt]{292.934pt}{0.400pt}}
\put(45,495){\makebox(0,0){$R_p(p_c)$}}
\put(828,23){\makebox(0,0){$L$}}
\put(1224,750){\makebox(0,0)[l]{$D = 3$}}
\put(1224,661){\makebox(0,0)[l]{$FBC$}}
\put(220.0,113.0){\rule[-0.200pt]{0.400pt}{184.048pt}}
\sbox{\plotpoint}{\rule[-0.500pt]{1.000pt}{1.000pt}}%
\put(503,877){\usebox{\plotpoint}}
\multiput(503,877)(8.866,-18.767){7}{\usebox{\plotpoint}}
\multiput(563,750)(10.901,-17.662){8}{\usebox{\plotpoint}}
\multiput(642,622)(14.887,-14.462){7}{\usebox{\plotpoint}}
\multiput(747,520)(18.356,-9.688){3}{\usebox{\plotpoint}}
\multiput(819,482)(19.121,-8.073){5}{\usebox{\plotpoint}}
\multiput(909,444)(19.853,-6.053){4}{\usebox{\plotpoint}}
\multiput(991,419)(20.179,-4.858){8}{\usebox{\plotpoint}}
\multiput(1153,380)(20.747,-0.584){4}{\usebox{\plotpoint}}
\multiput(1224,378)(20.735,-0.922){4}{\usebox{\plotpoint}}
\multiput(1314,374)(20.730,-1.020){6}{\usebox{\plotpoint}}
\put(1436,368){\usebox{\plotpoint}}
\put(1153,380){\usebox{\plotpoint}}
\multiput(1153,380)(20.416,3.738){4}{\usebox{\plotpoint}}
\multiput(1224,393)(18.491,9.427){3}{\usebox{\plotpoint}}
\multiput(1275,419)(13.944,15.374){3}{\usebox{\plotpoint}}
\multiput(1314,462)(11.558,17.239){5}{\usebox{\plotpoint}}
\put(1373,550){\usebox{\plotpoint}}
\put(1153,380){\usebox{\plotpoint}}
\multiput(1153,380)(20.050,-5.365){4}{\usebox{\plotpoint}}
\multiput(1224,361)(19.450,-7.246){3}{\usebox{\plotpoint}}
\multiput(1275,342)(17.474,-11.201){2}{\usebox{\plotpoint}}
\multiput(1314,317)(17.684,-10.866){5}{\usebox{\plotpoint}}
\put(1397,266){\usebox{\plotpoint}}
\sbox{\plotpoint}{\rule[-0.200pt]{0.400pt}{0.400pt}}%
\put(563,777){\makebox(0,0){$\bullet$}}
\put(672,584){\makebox(0,0){$\bullet$}}
\put(813,509){\makebox(0,0){$\bullet$}}
\put(912,458){\makebox(0,0){$\bullet$}}
\put(989,422){\makebox(0,0){$\bullet$}}
\put(1127,393){\makebox(0,0){$\bullet$}}
\put(1193,421){\makebox(0,0){$\bullet$}}
\put(1307,329){\makebox(0,0){$\bullet$}}
\put(563.0,768.0){\rule[-0.200pt]{0.400pt}{4.095pt}}
\put(553.0,768.0){\rule[-0.200pt]{4.818pt}{0.400pt}}
\put(553.0,785.0){\rule[-0.200pt]{4.818pt}{0.400pt}}
\put(672.0,574.0){\rule[-0.200pt]{0.400pt}{4.818pt}}
\put(662.0,574.0){\rule[-0.200pt]{4.818pt}{0.400pt}}
\put(662.0,594.0){\rule[-0.200pt]{4.818pt}{0.400pt}}
\put(813.0,495.0){\rule[-0.200pt]{0.400pt}{6.504pt}}
\put(803.0,495.0){\rule[-0.200pt]{4.818pt}{0.400pt}}
\put(803.0,522.0){\rule[-0.200pt]{4.818pt}{0.400pt}}
\put(912.0,444.0){\rule[-0.200pt]{0.400pt}{6.745pt}}
\put(902.0,444.0){\rule[-0.200pt]{4.818pt}{0.400pt}}
\put(902.0,472.0){\rule[-0.200pt]{4.818pt}{0.400pt}}
\put(989.0,412.0){\rule[-0.200pt]{0.400pt}{5.059pt}}
\put(979.0,412.0){\rule[-0.200pt]{4.818pt}{0.400pt}}
\put(979.0,433.0){\rule[-0.200pt]{4.818pt}{0.400pt}}
\put(1127.0,383.0){\rule[-0.200pt]{0.400pt}{4.818pt}}
\put(1117.0,383.0){\rule[-0.200pt]{4.818pt}{0.400pt}}
\put(1117.0,403.0){\rule[-0.200pt]{4.818pt}{0.400pt}}
\put(1193.0,405.0){\rule[-0.200pt]{0.400pt}{7.468pt}}
\put(1183.0,405.0){\rule[-0.200pt]{4.818pt}{0.400pt}}
\put(1183.0,436.0){\rule[-0.200pt]{4.818pt}{0.400pt}}
\put(1307.0,306.0){\rule[-0.200pt]{0.400pt}{11.322pt}}
\put(1297.0,306.0){\rule[-0.200pt]{4.818pt}{0.400pt}}
\put(1297.0,353.0){\rule[-0.200pt]{4.818pt}{0.400pt}}
\put(1252,415){\makebox(0,0){$\triangle$}}
\put(1307,450){\makebox(0,0){$\triangle$}}
\put(1252.0,390.0){\rule[-0.200pt]{0.400pt}{12.045pt}}
\put(1242.0,390.0){\rule[-0.200pt]{4.818pt}{0.400pt}}
\put(1242.0,440.0){\rule[-0.200pt]{4.818pt}{0.400pt}}
\put(1307.0,427.0){\rule[-0.200pt]{0.400pt}{11.081pt}}
\put(1297.0,427.0){\rule[-0.200pt]{4.818pt}{0.400pt}}
\put(1297.0,473.0){\rule[-0.200pt]{4.818pt}{0.400pt}}
\end{picture}

\vspace {0.1 cm}

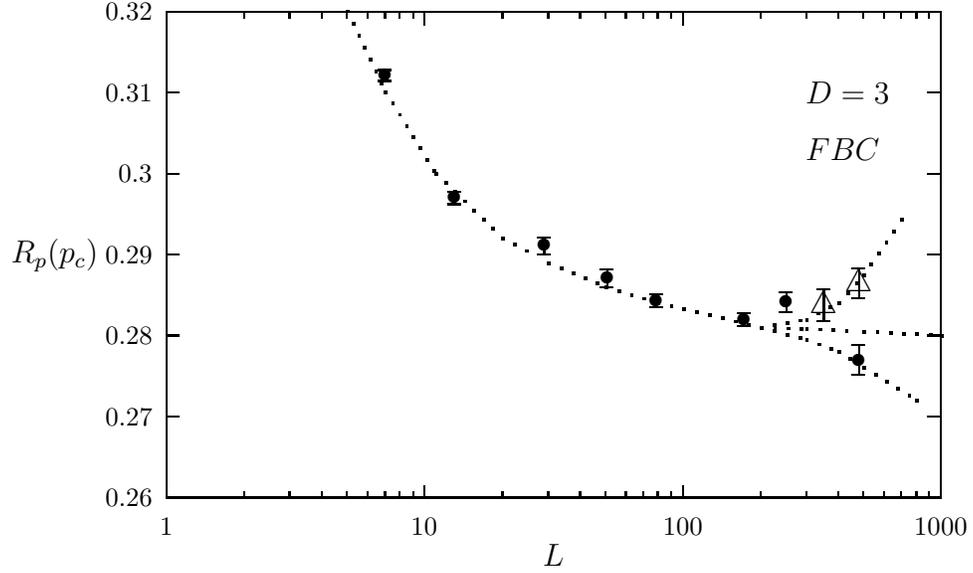
\noindent {\bf Fig. 3.} Size dependence of critical spanning probability
 in three dimensions for free boundary conditions. The symbols $\bullet$
represent $p_c = 0.3116$ and $\triangle$ represent $p_c$ = 0.31161. Dotted
lines are guides to the eye.

\vspace {0.1 cm}

% GNUPLOT: LaTeX picture
% Fig-4
\setlength{\unitlength}{0.240900pt}
\ifx\plotpoint\undefined\newsavebox{\plotpoint}\fi
\sbox{\plotpoint}{\rule[-0.200pt]{0.400pt}{0.400pt}}%
\begin{picture}(1500,900)(0,0)
\font\gnuplot=cmr10 at 10pt
\gnuplot
\sbox{\plotpoint}{\rule[-0.200pt]{0.400pt}{0.400pt}}%
\put(220.0,113.0){\rule[-0.200pt]{4.818pt}{0.400pt}}
\put(198,113){\makebox(0,0)[r]{0.38}}
\put(1416.0,113.0){\rule[-0.200pt]{4.818pt}{0.400pt}}
\put(220.0,209.0){\rule[-0.200pt]{4.818pt}{0.400pt}}
\put(198,209){\makebox(0,0)[r]{0.385}}
\put(1416.0,209.0){\rule[-0.200pt]{4.818pt}{0.400pt}}
\put(220.0,304.0){\rule[-0.200pt]{4.818pt}{0.400pt}}
\put(198,304){\makebox(0,0)[r]{0.39}}
\put(1416.0,304.0){\rule[-0.200pt]{4.818pt}{0.400pt}}
\put(220.0,400.0){\rule[-0.200pt]{4.818pt}{0.400pt}}
\put(198,400){\makebox(0,0)[r]{0.395}}
\put(1416.0,400.0){\rule[-0.200pt]{4.818pt}{0.400pt}}
\put(220.0,495.0){\rule[-0.200pt]{4.818pt}{0.400pt}}
\put(198,495){\makebox(0,0)[r]{0.4}}
\put(1416.0,495.0){\rule[-0.200pt]{4.818pt}{0.400pt}}
\put(220.0,591.0){\rule[-0.200pt]{4.818pt}{0.400pt}}
\put(198,591){\makebox(0,0)[r]{0.405}}
\put(1416.0,591.0){\rule[-0.200pt]{4.818pt}{0.400pt}}
\put(220.0,686.0){\rule[-0.200pt]{4.818pt}{0.400pt}}
\put(198,686){\makebox(0,0)[r]{0.41}}
\put(1416.0,686.0){\rule[-0.200pt]{4.818pt}{0.400pt}}
\put(220.0,782.0){\rule[-0.200pt]{4.818pt}{0.400pt}}
\put(198,782){\makebox(0,0)[r]{0.415}}
\put(1416.0,782.0){\rule[-0.200pt]{4.818pt}{0.400pt}}
\put(220.0,877.0){\rule[-0.200pt]{4.818pt}{0.400pt}}
\put(198,877){\makebox(0,0)[r]{0.42}}
\put(1416.0,877.0){\rule[-0.200pt]{4.818pt}{0.400pt}}
\put(220.0,113.0){\rule[-0.200pt]{0.400pt}{4.818pt}}
\put(220,68){\makebox(0,0){1}}
\put(220.0,857.0){\rule[-0.200pt]{0.400pt}{4.818pt}}
\put(342.0,113.0){\rule[-0.200pt]{0.400pt}{2.409pt}}
\put(342.0,867.0){\rule[-0.200pt]{0.400pt}{2.409pt}}
\put(413.0,113.0){\rule[-0.200pt]{0.400pt}{2.409pt}}
\put(413.0,867.0){\rule[-0.200pt]{0.400pt}{2.409pt}}
\put(464.0,113.0){\rule[-0.200pt]{0.400pt}{2.409pt}}
\put(464.0,867.0){\rule[-0.200pt]{0.400pt}{2.409pt}}
\put(503.0,113.0){\rule[-0.200pt]{0.400pt}{2.409pt}}
\put(503.0,867.0){\rule[-0.200pt]{0.400pt}{2.409pt}}
\put(535.0,113.0){\rule[-0.200pt]{0.400pt}{2.409pt}}
\put(535.0,867.0){\rule[-0.200pt]{0.400pt}{2.409pt}}
\put(563.0,113.0){\rule[-0.200pt]{0.400pt}{2.409pt}}
\put(563.0,867.0){\rule[-0.200pt]{0.400pt}{2.409pt}}
\put(586.0,113.0){\rule[-0.200pt]{0.400pt}{2.409pt}}
\put(586.0,867.0){\rule[-0.200pt]{0.400pt}{2.409pt}}
\put(607.0,113.0){\rule[-0.200pt]{0.400pt}{2.409pt}}
\put(607.0,867.0){\rule[-0.200pt]{0.400pt}{2.409pt}}
\put(625.0,113.0){\rule[-0.200pt]{0.400pt}{4.818pt}}
\put(625,68){\makebox(0,0){10}}
\put(625.0,857.0){\rule[-0.200pt]{0.400pt}{4.818pt}}
\put(747.0,113.0){\rule[-0.200pt]{0.400pt}{2.409pt}}
\put(747.0,867.0){\rule[-0.200pt]{0.400pt}{2.409pt}}
\put(819.0,113.0){\rule[-0.200pt]{0.400pt}{2.409pt}}
\put(819.0,867.0){\rule[-0.200pt]{0.400pt}{2.409pt}}
\put(869.0,113.0){\rule[-0.200pt]{0.400pt}{2.409pt}}
\put(869.0,867.0){\rule[-0.200pt]{0.400pt}{2.409pt}}
\put(909.0,113.0){\rule[-0.200pt]{0.400pt}{2.409pt}}
\put(909.0,867.0){\rule[-0.200pt]{0.400pt}{2.409pt}}
\put(941.0,113.0){\rule[-0.200pt]{0.400pt}{2.409pt}}
\put(941.0,867.0){\rule[-0.200pt]{0.400pt}{2.409pt}}
\put(968.0,113.0){\rule[-0.200pt]{0.400pt}{2.409pt}}
\put(968.0,867.0){\rule[-0.200pt]{0.400pt}{2.409pt}}
\put(991.0,113.0){\rule[-0.200pt]{0.400pt}{2.409pt}}
\put(991.0,867.0){\rule[-0.200pt]{0.400pt}{2.409pt}}
\put(1012.0,113.0){\rule[-0.200pt]{0.400pt}{2.409pt}}
\put(1012.0,867.0){\rule[-0.200pt]{0.400pt}{2.409pt}}
\put(1031.0,113.0){\rule[-0.200pt]{0.400pt}{4.818pt}}
\put(1031,68){\makebox(0,0){100}}
\put(1031.0,857.0){\rule[-0.200pt]{0.400pt}{4.818pt}}
\put(1153.0,113.0){\rule[-0.200pt]{0.400pt}{2.409pt}}
\put(1153.0,867.0){\rule[-0.200pt]{0.400pt}{2.409pt}}
\put(1224.0,113.0){\rule[-0.200pt]{0.400pt}{2.409pt}}
\put(1224.0,867.0){\rule[-0.200pt]{0.400pt}{2.409pt}}
\put(1275.0,113.0){\rule[-0.200pt]{0.400pt}{2.409pt}}
\put(1275.0,867.0){\rule[-0.200pt]{0.400pt}{2.409pt}}
\put(1314.0,113.0){\rule[-0.200pt]{0.400pt}{2.409pt}}
\put(1314.0,867.0){\rule[-0.200pt]{0.400pt}{2.409pt}}
\put(1346.0,113.0){\rule[-0.200pt]{0.400pt}{2.409pt}}
\put(1346.0,867.0){\rule[-0.200pt]{0.400pt}{2.409pt}}
\put(1373.0,113.0){\rule[-0.200pt]{0.400pt}{2.409pt}}
\put(1373.0,867.0){\rule[-0.200pt]{0.400pt}{2.409pt}}
\put(1397.0,113.0){\rule[-0.200pt]{0.400pt}{2.409pt}}
\put(1397.0,867.0){\rule[-0.200pt]{0.400pt}{2.409pt}}
\put(1417.0,113.0){\rule[-0.200pt]{0.400pt}{2.409pt}}
\put(1417.0,867.0){\rule[-0.200pt]{0.400pt}{2.409pt}}
\put(1436.0,113.0){\rule[-0.200pt]{0.400pt}{4.818pt}}
\put(1436,68){\makebox(0,0){1000}}
\put(1436.0,857.0){\rule[-0.200pt]{0.400pt}{4.818pt}}
\put(220.0,113.0){\rule[-0.200pt]{292.934pt}{0.400pt}}
\put(1436.0,113.0){\rule[-0.200pt]{0.400pt}{184.048pt}}
\put(220.0,877.0){\rule[-0.200pt]{292.934pt}{0.400pt}}
\put(45,495){\makebox(0,0){$R_p(p_c)$}}
\put(828,23){\makebox(0,0){$L$}}
\put(413,781){\makebox(0,0)[l]{$D = 3$}}
\put(413,686){\makebox(0,0)[l]{$MBC$}}
\put(220.0,113.0){\rule[-0.200pt]{0.400pt}{184.048pt}}
\sbox{\plotpoint}{\rule[-0.500pt]{1.000pt}{1.000pt}}%
\put(1436,686){\usebox{\plotpoint}}
\multiput(1436,686)(-20.753,-0.340){6}{\usebox{\plotpoint}}
\multiput(1314,684)(-20.674,-1.838){5}{\usebox{\plotpoint}}
\multiput(1224,676)(-20.656,-2.033){9}{\usebox{\plotpoint}}
\multiput(1031,657)(-20.365,-4.006){6}{\usebox{\plotpoint}}
\multiput(909,633)(-19.624,-6.759){5}{\usebox{\plotpoint}}
\multiput(819,602)(-18.144,-10.080){4}{\usebox{\plotpoint}}
\multiput(747,562)(-17.004,-11.902){3}{\usebox{\plotpoint}}
\multiput(697,527)(-17.601,-11.000){4}{\usebox{\plotpoint}}
\multiput(625,482)(-15.651,-13.632){4}{\usebox{\plotpoint}}
\multiput(563,428)(-15.997,-13.224){9}{\usebox{\plotpoint}}
\put(413,304){\usebox{\plotpoint}}
\put(1127,667){\usebox{\plotpoint}}
\multiput(1127,667)(17.895,10.515){6}{\usebox{\plotpoint}}
\multiput(1224,724)(16.643,12.401){3}{\usebox{\plotpoint}}
\multiput(1275,762)(14.676,14.676){3}{\usebox{\plotpoint}}
\multiput(1314,801)(13.369,15.876){2}{\usebox{\plotpoint}}
\put(1346,839){\usebox{\plotpoint}}
\put(1127,667){\usebox{\plotpoint}}
\multiput(1127,667)(20.368,-3.990){5}{\usebox{\plotpoint}}
\multiput(1224,648)(19.450,-7.246){3}{\usebox{\plotpoint}}
\multiput(1275,629)(18.659,-9.090){2}{\usebox{\plotpoint}}
\multiput(1314,610)(17.847,-10.596){2}{\usebox{\plotpoint}}
\put(1346,591){\usebox{\plotpoint}}
\sbox{\plotpoint}{\rule[-0.200pt]{0.400pt}{0.400pt}}%
\put(563,535){\makebox(0,0){$\bullet$}}
\put(672,508){\makebox(0,0){$\bullet$}}
\put(813,600){\makebox(0,0){$\bullet$}}
\put(912,652){\makebox(0,0){$\bullet$}}
\put(989,666){\makebox(0,0){$\bullet$}}
\put(1127,668){\makebox(0,0){$\bullet$}}
\put(1193,740){\makebox(0,0){$\bullet$}}
\put(1307,619){\makebox(0,0){$\bullet$}}
\put(563.0,513.0){\rule[-0.200pt]{0.400pt}{10.840pt}}
\put(553.0,513.0){\rule[-0.200pt]{4.818pt}{0.400pt}}
\put(553.0,558.0){\rule[-0.200pt]{4.818pt}{0.400pt}}
\put(672.0,492.0){\rule[-0.200pt]{0.400pt}{7.709pt}}
\put(662.0,492.0){\rule[-0.200pt]{4.818pt}{0.400pt}}
\put(662.0,524.0){\rule[-0.200pt]{4.818pt}{0.400pt}}
\put(813.0,585.0){\rule[-0.200pt]{0.400pt}{7.227pt}}
\put(803.0,585.0){\rule[-0.200pt]{4.818pt}{0.400pt}}
\put(803.0,615.0){\rule[-0.200pt]{4.818pt}{0.400pt}}
\put(912.0,628.0){\rule[-0.200pt]{0.400pt}{11.563pt}}
\put(902.0,628.0){\rule[-0.200pt]{4.818pt}{0.400pt}}
\put(902.0,676.0){\rule[-0.200pt]{4.818pt}{0.400pt}}
\put(989.0,649.0){\rule[-0.200pt]{0.400pt}{8.191pt}}
\put(979.0,649.0){\rule[-0.200pt]{4.818pt}{0.400pt}}
\put(979.0,683.0){\rule[-0.200pt]{4.818pt}{0.400pt}}
\put(1127.0,649.0){\rule[-0.200pt]{0.400pt}{9.395pt}}
\put(1117.0,649.0){\rule[-0.200pt]{4.818pt}{0.400pt}}
\put(1117.0,688.0){\rule[-0.200pt]{4.818pt}{0.400pt}}
\put(1193.0,714.0){\rule[-0.200pt]{0.400pt}{12.527pt}}
\put(1183.0,714.0){\rule[-0.200pt]{4.818pt}{0.400pt}}
\put(1183.0,766.0){\rule[-0.200pt]{4.818pt}{0.400pt}}
\put(1307.0,580.0){\rule[-0.200pt]{0.400pt}{19.031pt}}
\put(1297.0,580.0){\rule[-0.200pt]{4.818pt}{0.400pt}}
\put(1297.0,659.0){\rule[-0.200pt]{4.818pt}{0.400pt}}
\put(1252,757){\makebox(0,0){$\triangle$}}
\put(1307,797){\makebox(0,0){$\triangle$}}
\put(1252.0,717.0){\rule[-0.200pt]{0.400pt}{19.272pt}}
\put(1242.0,717.0){\rule[-0.200pt]{4.818pt}{0.400pt}}
\put(1242.0,797.0){\rule[-0.200pt]{4.818pt}{0.400pt}}
\put(1307.0,755.0){\rule[-0.200pt]{0.400pt}{19.995pt}}
\put(1297.0,755.0){\rule[-0.200pt]{4.818pt}{0.400pt}}
\put(1297.0,838.0){\rule[-0.200pt]{4.818pt}{0.400pt}}
\end{picture}

\vspace {0.1 cm}

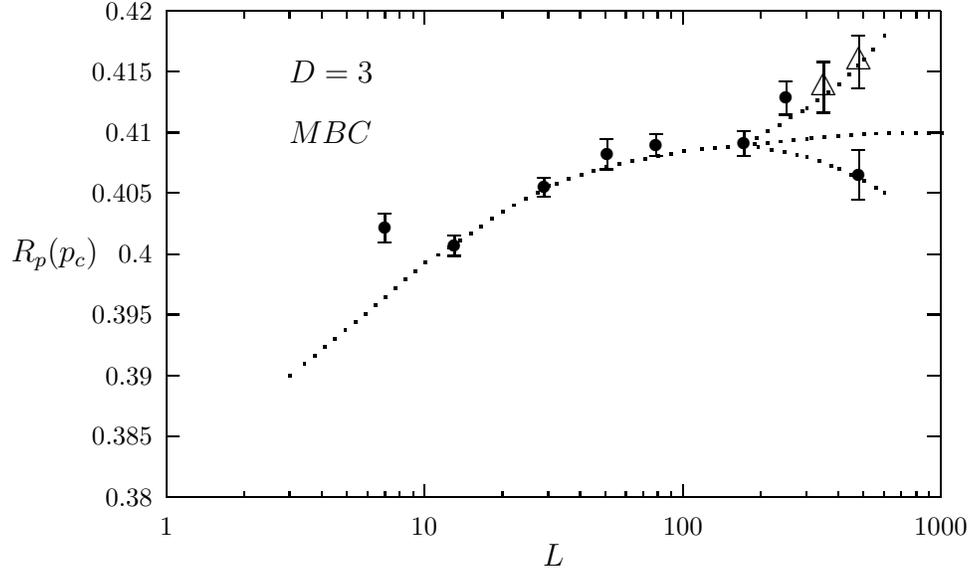
\noindent {\bf Fig. 4.} As Fig. 3, for mixed boundary conditions.

\newpage

% GNUPLOT: LaTeX picture
% Fig-5
\setlength{\unitlength}{0.240900pt}
\ifx\plotpoint\undefined\newsavebox{\plotpoint}\fi
\sbox{\plotpoint}{\rule[-0.200pt]{0.400pt}{0.400pt}}%
\begin{picture}(1500,900)(0,0)
\font\gnuplot=cmr10 at 10pt
\gnuplot
\sbox{\plotpoint}{\rule[-0.200pt]{0.400pt}{0.400pt}}%
\put(220.0,113.0){\rule[-0.200pt]{4.818pt}{0.400pt}}
\put(198,113){\makebox(0,0)[r]{0.48}}
\put(1416.0,113.0){\rule[-0.200pt]{4.818pt}{0.400pt}}
\put(220.0,198.0){\rule[-0.200pt]{4.818pt}{0.400pt}}
\put(198,198){\makebox(0,0)[r]{0.485}}
\put(1416.0,198.0){\rule[-0.200pt]{4.818pt}{0.400pt}}
\put(220.0,283.0){\rule[-0.200pt]{4.818pt}{0.400pt}}
\put(198,283){\makebox(0,0)[r]{0.49}}
\put(1416.0,283.0){\rule[-0.200pt]{4.818pt}{0.400pt}}
\put(220.0,368.0){\rule[-0.200pt]{4.818pt}{0.400pt}}
\put(198,368){\makebox(0,0)[r]{0.495}}
\put(1416.0,368.0){\rule[-0.200pt]{4.818pt}{0.400pt}}
\put(220.0,453.0){\rule[-0.200pt]{4.818pt}{0.400pt}}
\put(198,453){\makebox(0,0)[r]{0.5}}
\put(1416.0,453.0){\rule[-0.200pt]{4.818pt}{0.400pt}}
\put(220.0,537.0){\rule[-0.200pt]{4.818pt}{0.400pt}}
\put(198,537){\makebox(0,0)[r]{0.505}}
\put(1416.0,537.0){\rule[-0.200pt]{4.818pt}{0.400pt}}
\put(220.0,622.0){\rule[-0.200pt]{4.818pt}{0.400pt}}
\put(198,622){\makebox(0,0)[r]{0.51}}
\put(1416.0,622.0){\rule[-0.200pt]{4.818pt}{0.400pt}}
\put(220.0,707.0){\rule[-0.200pt]{4.818pt}{0.400pt}}
\put(198,707){\makebox(0,0)[r]{0.515}}
\put(1416.0,707.0){\rule[-0.200pt]{4.818pt}{0.400pt}}
\put(220.0,792.0){\rule[-0.200pt]{4.818pt}{0.400pt}}
\put(198,792){\makebox(0,0)[r]{0.52}}
\put(1416.0,792.0){\rule[-0.200pt]{4.818pt}{0.400pt}}
\put(220.0,877.0){\rule[-0.200pt]{4.818pt}{0.400pt}}
\put(198,877){\makebox(0,0)[r]{0.525}}
\put(1416.0,877.0){\rule[-0.200pt]{4.818pt}{0.400pt}}
\put(220.0,113.0){\rule[-0.200pt]{0.400pt}{4.818pt}}
\put(220,68){\makebox(0,0){1}}
\put(220.0,857.0){\rule[-0.200pt]{0.400pt}{4.818pt}}
\put(342.0,113.0){\rule[-0.200pt]{0.400pt}{2.409pt}}
\put(342.0,867.0){\rule[-0.200pt]{0.400pt}{2.409pt}}
\put(413.0,113.0){\rule[-0.200pt]{0.400pt}{2.409pt}}
\put(413.0,867.0){\rule[-0.200pt]{0.400pt}{2.409pt}}
\put(464.0,113.0){\rule[-0.200pt]{0.400pt}{2.409pt}}
\put(464.0,867.0){\rule[-0.200pt]{0.400pt}{2.409pt}}
\put(503.0,113.0){\rule[-0.200pt]{0.400pt}{2.409pt}}
\put(503.0,867.0){\rule[-0.200pt]{0.400pt}{2.409pt}}
\put(535.0,113.0){\rule[-0.200pt]{0.400pt}{2.409pt}}
\put(535.0,867.0){\rule[-0.200pt]{0.400pt}{2.409pt}}
\put(563.0,113.0){\rule[-0.200pt]{0.400pt}{2.409pt}}
\put(563.0,867.0){\rule[-0.200pt]{0.400pt}{2.409pt}}
\put(586.0,113.0){\rule[-0.200pt]{0.400pt}{2.409pt}}
\put(586.0,867.0){\rule[-0.200pt]{0.400pt}{2.409pt}}
\put(607.0,113.0){\rule[-0.200pt]{0.400pt}{2.409pt}}
\put(607.0,867.0){\rule[-0.200pt]{0.400pt}{2.409pt}}
\put(625.0,113.0){\rule[-0.200pt]{0.400pt}{4.818pt}}
\put(625,68){\makebox(0,0){10}}
\put(625.0,857.0){\rule[-0.200pt]{0.400pt}{4.818pt}}
\put(747.0,113.0){\rule[-0.200pt]{0.400pt}{2.409pt}}
\put(747.0,867.0){\rule[-0.200pt]{0.400pt}{2.409pt}}
\put(819.0,113.0){\rule[-0.200pt]{0.400pt}{2.409pt}}
\put(819.0,867.0){\rule[-0.200pt]{0.400pt}{2.409pt}}
\put(869.0,113.0){\rule[-0.200pt]{0.400pt}{2.409pt}}
\put(869.0,867.0){\rule[-0.200pt]{0.400pt}{2.409pt}}
\put(909.0,113.0){\rule[-0.200pt]{0.400pt}{2.409pt}}
\put(909.0,867.0){\rule[-0.200pt]{0.400pt}{2.409pt}}
\put(941.0,113.0){\rule[-0.200pt]{0.400pt}{2.409pt}}
\put(941.0,867.0){\rule[-0.200pt]{0.400pt}{2.409pt}}
\put(968.0,113.0){\rule[-0.200pt]{0.400pt}{2.409pt}}
\put(968.0,867.0){\rule[-0.200pt]{0.400pt}{2.409pt}}
\put(991.0,113.0){\rule[-0.200pt]{0.400pt}{2.409pt}}
\put(991.0,867.0){\rule[-0.200pt]{0.400pt}{2.409pt}}
\put(1012.0,113.0){\rule[-0.200pt]{0.400pt}{2.409pt}}
\put(1012.0,867.0){\rule[-0.200pt]{0.400pt}{2.409pt}}
\put(1031.0,113.0){\rule[-0.200pt]{0.400pt}{4.818pt}}
\put(1031,68){\makebox(0,0){100}}
\put(1031.0,857.0){\rule[-0.200pt]{0.400pt}{4.818pt}}
\put(1153.0,113.0){\rule[-0.200pt]{0.400pt}{2.409pt}}
\put(1153.0,867.0){\rule[-0.200pt]{0.400pt}{2.409pt}}
\put(1224.0,113.0){\rule[-0.200pt]{0.400pt}{2.409pt}}
\put(1224.0,867.0){\rule[-0.200pt]{0.400pt}{2.409pt}}
\put(1275.0,113.0){\rule[-0.200pt]{0.400pt}{2.409pt}}
\put(1275.0,867.0){\rule[-0.200pt]{0.400pt}{2.409pt}}
\put(1314.0,113.0){\rule[-0.200pt]{0.400pt}{2.409pt}}
\put(1314.0,867.0){\rule[-0.200pt]{0.400pt}{2.409pt}}
\put(1346.0,113.0){\rule[-0.200pt]{0.400pt}{2.409pt}}
\put(1346.0,867.0){\rule[-0.200pt]{0.400pt}{2.409pt}}
\put(1373.0,113.0){\rule[-0.200pt]{0.400pt}{2.409pt}}
\put(1373.0,867.0){\rule[-0.200pt]{0.400pt}{2.409pt}}
\put(1397.0,113.0){\rule[-0.200pt]{0.400pt}{2.409pt}}
\put(1397.0,867.0){\rule[-0.200pt]{0.400pt}{2.409pt}}
\put(1417.0,113.0){\rule[-0.200pt]{0.400pt}{2.409pt}}
\put(1417.0,867.0){\rule[-0.200pt]{0.400pt}{2.409pt}}
\put(1436.0,113.0){\rule[-0.200pt]{0.400pt}{4.818pt}}
\put(1436,68){\makebox(0,0){1000}}
\put(1436.0,857.0){\rule[-0.200pt]{0.400pt}{4.818pt}}
\put(220.0,113.0){\rule[-0.200pt]{292.934pt}{0.400pt}}
\put(1436.0,113.0){\rule[-0.200pt]{0.400pt}{184.048pt}}
\put(220.0,877.0){\rule[-0.200pt]{292.934pt}{0.400pt}}
\put(45,495){\makebox(0,0){$R_p(p_c)$}}
\put(828,23){\makebox(0,0){$L$}}
\put(413,792){\makebox(0,0)[l]{$D = 3$}}
\put(413,707){\makebox(0,0)[l]{$HBC$}}
\put(220.0,113.0){\rule[-0.200pt]{0.400pt}{184.048pt}}
\sbox{\plotpoint}{\rule[-0.500pt]{1.000pt}{1.000pt}}%
\put(1436,670){\usebox{\plotpoint}}
\multiput(1436,670)(-20.756,0.000){22}{\usebox{\plotpoint}}
\multiput(989,670)(-20.586,-2.647){3}{\usebox{\plotpoint}}
\multiput(919,661)(-17.482,-11.188){6}{\usebox{\plotpoint}}
\multiput(819,597)(-11.212,-17.467){13}{\usebox{\plotpoint}}
\multiput(672,368)(-11.203,-17.472){10}{\usebox{\plotpoint}}
\multiput(563,198)(-11.969,-16.957){5}{\usebox{\plotpoint}}
\put(503,113){\usebox{\plotpoint}}
\put(1193,670){\usebox{\plotpoint}}
\multiput(1193,670)(20.339,4.137){3}{\usebox{\plotpoint}}
\multiput(1252,682)(12.168,16.814){5}{\usebox{\plotpoint}}
\multiput(1307,758)(7.413,19.387){5}{\usebox{\plotpoint}}
\put(1346,860){\usebox{\plotpoint}}
\put(1193,670){\usebox{\plotpoint}}
\multiput(1193,670)(13.833,-15.474){5}{\usebox{\plotpoint}}
\multiput(1252,604)(11.370,-17.364){5}{\usebox{\plotpoint}}
\multiput(1307,520)(7.476,-19.362){5}{\usebox{\plotpoint}}
\put(1346,419){\usebox{\plotpoint}}
\sbox{\plotpoint}{\rule[-0.200pt]{0.400pt}{0.400pt}}%
\put(563,198){\makebox(0,0){$\bullet$}}
\put(672,365){\makebox(0,0){$\bullet$}}
\put(813,577){\makebox(0,0){$\bullet$}}
\put(912,676){\makebox(0,0){$\bullet$}}
\put(989,685){\makebox(0,0){$\bullet$}}
\put(1127,664){\makebox(0,0){$\bullet$}}
\put(1193,661){\makebox(0,0){$\bullet$}}
\put(1307,523){\makebox(0,0){$\bullet$}}
\put(563.0,181.0){\rule[-0.200pt]{0.400pt}{8.191pt}}
\put(553.0,181.0){\rule[-0.200pt]{4.818pt}{0.400pt}}
\put(553.0,215.0){\rule[-0.200pt]{4.818pt}{0.400pt}}
\put(672.0,345.0){\rule[-0.200pt]{0.400pt}{9.395pt}}
\put(662.0,345.0){\rule[-0.200pt]{4.818pt}{0.400pt}}
\put(662.0,384.0){\rule[-0.200pt]{4.818pt}{0.400pt}}
\put(813.0,560.0){\rule[-0.200pt]{0.400pt}{7.950pt}}
\put(803.0,560.0){\rule[-0.200pt]{4.818pt}{0.400pt}}
\put(803.0,593.0){\rule[-0.200pt]{4.818pt}{0.400pt}}
\put(912.0,658.0){\rule[-0.200pt]{0.400pt}{8.672pt}}
\put(902.0,658.0){\rule[-0.200pt]{4.818pt}{0.400pt}}
\put(902.0,694.0){\rule[-0.200pt]{4.818pt}{0.400pt}}
\put(989.0,667.0){\rule[-0.200pt]{0.400pt}{8.431pt}}
\put(979.0,667.0){\rule[-0.200pt]{4.818pt}{0.400pt}}
\put(979.0,702.0){\rule[-0.200pt]{4.818pt}{0.400pt}}
\put(1127.0,647.0){\rule[-0.200pt]{0.400pt}{8.191pt}}
\put(1117.0,647.0){\rule[-0.200pt]{4.818pt}{0.400pt}}
\put(1117.0,681.0){\rule[-0.200pt]{4.818pt}{0.400pt}}
\put(1193.0,637.0){\rule[-0.200pt]{0.400pt}{11.804pt}}
\put(1183.0,637.0){\rule[-0.200pt]{4.818pt}{0.400pt}}
\put(1183.0,686.0){\rule[-0.200pt]{4.818pt}{0.400pt}}
\put(1307.0,485.0){\rule[-0.200pt]{0.400pt}{18.308pt}}
\put(1297.0,485.0){\rule[-0.200pt]{4.818pt}{0.400pt}}
\put(1297.0,561.0){\rule[-0.200pt]{4.818pt}{0.400pt}}
\put(1252,668){\makebox(0,0){$\triangle$}}
\put(1307,781){\makebox(0,0){$\triangle$}}
\put(1252.0,638.0){\rule[-0.200pt]{0.400pt}{14.213pt}}
\put(1242.0,638.0){\rule[-0.200pt]{4.818pt}{0.400pt}}
\put(1242.0,697.0){\rule[-0.200pt]{4.818pt}{0.400pt}}
\put(1307.0,743.0){\rule[-0.200pt]{0.400pt}{18.067pt}}
\put(1297.0,743.0){\rule[-0.200pt]{4.818pt}{0.400pt}}
\put(1297.0,818.0){\rule[-0.200pt]{4.818pt}{0.400pt}}
\end{picture}

\bigskip

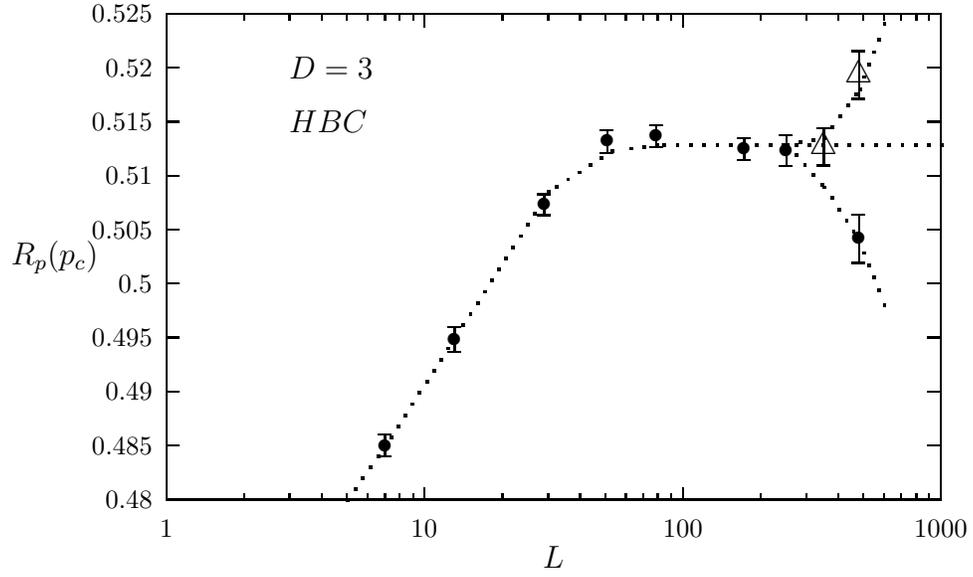
\noindent {\bf Fig. 5.} As Fig. 3, for helical boundary conditions.

\end{document}